\documentclass[12pt]{article}
\usepackage{natbib}
\usepackage[english]{babel}
\usepackage{amsmath}
\usepackage{amssymb}
\usepackage{amsthm}
\usepackage{mathtools}
\usepackage{physics}
\usepackage{epsfig}
\usepackage{graphics}
\usepackage{bm}
\usepackage{tikz}
\usetikzlibrary{arrows,positioning,shapes.geometric,decorations.markings}
\tikzstyle{squarenode}=[rectangle,draw]
\tikzstyle{littlenode}=[circle,draw,minimum size=0.5cm,font=\small]
\tikzstyle{bigellipse}=[ellipse,draw,x radius=10cm, y radius=5cm]
\tikzset{negate/.style={
            decoration={markings,
            mark= at position 0.5 with {
                  \node[transform shape] (tempnode) {$\Big\Vert$};
                  }
              },
              postaction={decorate}
}
}
\usepackage{xcolor}
\usepackage{tabularx}
\usepackage{multirow}
\usepackage{booktabs}
\usepackage{rotating}
\usepackage{caption}
\usepackage{floatrow}
\usepackage{graphicx} 
\usepackage{graphics} 
\usepackage{subfig}
\usepackage{pgfplots}
\usepackage{authblk}
\usepackage{etoolbox}

\usepackage{soul}

\newcommand{\ind}{\mbox{$\perp\!\!\!\perp$}}

\newcommand{\logit}{\mbox{logit}}

\pgfplotsset{compat=1.5}
\RequirePackage{hyperref}

\hypersetup{
	colorlinks=true,
	linkcolor=black,
	citecolor=black,
	filecolor=black,
	urlcolor=black,
}
\def\ind{\perp\!\!\!\perp}

\theoremstyle{plain}

\title{Exact mediation analysis for ordinal outcome and binary mediator
\thanks{An extended version is published in {\it Epidemiology}, Vol. 33, Num. 6, November 2022. }}
\date{}

\author[1]{Elena Stanghellini}
\affil[1]{Department of Economics, University of Perugia, Italy}
\author[2]{Maria Kateri}
\affil[2]{Institute of Statistics, RWTH Aachen University, Germany}

\begin{document}
\maketitle

\section*{Abstract}

With reference to a single mediator context, this brief report 
presents a model-based strategy to estimate counterfactual direct and indirect effects when the response variable is ordinal and the mediator is binary. Postulating
a logistic regression model for the mediator and a cumulative logit model for the 
outcome, we present the exact parametric formulation of the causal effects, thereby extending previous work that only contained approximated results. The identification conditions are equivalent to the ones already 
established in the literature. The effects can be estimated by making use of standard statistical software and standard errors can be computed via  a
bootstrap algorithm. To make the methodology accessible, routines to implement the proposal  in R are presented in the eAppendix. We also derive the natural effect model coherent with the postulated data-generating mechanism. 

\vspace{0.1cm}

{\bf keywords}: binary mediator, causal effects, mediation, natural effect model, ordinal outcome

%\vspace{0.1cm}
\section*{Introduction}

Many epidemiologic problems involve the quantification of the causal effect of a treatment on an outcome and the decomposition of this effect into the direct and indirect one, this second due to the presence of a possible mediator. A mediator is a variable that is a response to the treatment and that in turn influences the outcome.

Let $X$ be a treatment of interest, $M$ be the mediator, and $Y$ the outcome of interest. We assume that the data-generating process, possibly after conditioning on a set $C$ of covariates, is as described in the Figure. Mediation analysis involves the definition and estimation of effects 
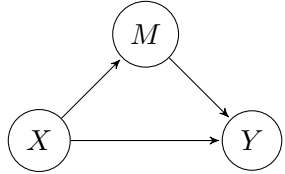
\begin{figure}[h] 
\centering{
\begin{tikzpicture}[scale=0.4,auto,->,>=stealth',shorten >=1pt,node distance=2cm] 
\node[littlenode] (M) {$M$};
\node[littlenode] (X) [below left of=M] {$X$};
\node[littlenode] (Y) [below right of=M] {$Y$};
\draw[->] (M) --node {} (Y) ; \draw[->] (X) --node {} (M); \draw[->] (X) --node {} (Y);
\end{tikzpicture}
\caption*{Figure: Data-generating process for one outcome $Y$ and one mediator $M$}\label{fig:dag1}}
\end{figure}
on the outcome $Y$ when, possibly contrary to fact, the value of $X$ is set to $x$ versus a baseline value $x^*$; see \cite{pearl2012mediation} $^1$ or \cite{VanderWeele2015book} $^2$.   

We assume that the outcome of interest is a categorical random variable with levels that can be ordered. Differently from what presented in \cite{VdWVanZhang2016}$^3$, we  do not make the assumption that the response  $Y$ is highly unbalanced, with one reference category having a high conditional probability (possibly higher than 0.90). Furthermore, we focus on a situation where the mediator $M$ is binary. We  present the closed form of the counterfactual effects for the described context and detail how to perform inference when a random sample drawn from the population is available. Standard statistical software may be used for the implementation of the proposed methodology. This paper extends the derivation in  \cite{doretti2018exact}$^4$, where the binary $M$ - binary $Y$ case is presented, and fills the current gap on existing parametric methods for causal inference and mediation to cover the described situation. Since the closed form of the effects is presented, the derivations also allow formulation of the natural effect model coherently with the postulated data-generating process.

\section*{Concepts and definitions}

Let $M(x)$, $Y(x)$  be, in order, the value that $M$ and $Y$ would take if $X$ were 
set to $x$. Let $Y(x,m)$ be the value of $Y$ if $X$ were set to $x$ and $M$ were set to $m$. Finally let $Y({\color{blue}x},M(x^*))$  be the value that $Y$ would take were $X$ set to $x$ and $M$ set to $M(x^*)$. i.e. the value that it would have naturally attained if $X$ were set to $x^*$. These values are called potential outcomes. For more details see \cite{VanderWeele2015book}$^2$, Ch. 1-2.

%\st{ We here make the so-called consistency and composition assumptions. Consistency states that, in the subgroup of units with $X=x$, the observed variables $Y$ and  $M$ equal the potential outcome variables $Y(x)$ and $M(x)$ respectively and that for units with $X=x$ and $M=m$ the observed value of $Y$ equals the potential outcome $Y(x,m)$. Composition requires that $Y(x)=Y(x,M(x))$, i.e. that the potential outcome associated to the intervention $X=x$  would be equal to the potential outcome
%associated to setting $X$ to $x$ and the mediator to $M(x)$, which is the value it would have naturally attained under $X=x$ either at individual or at population level; see Kuha for a discussion. Following Pearl2001, we here focus on contrasts at population level. }

Causal effects involve contrasts on the potential outcome for $X$ set to $x$ or $x^*$, with $x^*$ a baseline level. 
\cite{VdWVanZhang2016}$^3$ introduced the causal effects for an ordinal outcome on a cumulative odds ratio scale. 
To identify the causal effects, further conditions are necessary as in \cite{VdWVanZhang2016}$^3$. A detailed presentation of these causal effects and their identification conditions is in eAppendix A.1. We here summarize them by assuming that, possibly after conditioning on observed covariates $C$, the DAG in the Figure describes the data-generating mechanism and no unobserved confounders exist; see \cite{PearlCausalityBook2009}$^6$, Ch. 7.  We further assume a parametric formulation of the data-generating process according to  models \eqref{eq:lrw} and \eqref{eq:lry}, provided in next section.

\section*{Exact parametric formulation of natural effects}

%\section{The modeling strategy}

We assume that 
\begin{equation}\label{eq:lrw}
\logit P(M=1\mid X=x, C=c)= \gamma_{0} +\gamma^Xx+\gamma^Cc
\end{equation}
and
\begin{equation}\label{eq:lry}
\logit P(Y\leq j\mid X=x,M=m,C=c) = \alpha_j-(\beta^{X}x+\beta^{M}m+\beta^{XM}xm+ \beta^Cc), \,\, j=1,\ldots, J-1
\end{equation}
\noindent where
\[
\logit P(Y\leq j\mid X=x,M=m,C=c) = \log \frac{P(Y\leq j \mid X=x,M=m,C=c)}{P(Y> j \mid X=x,M=m,C=c)}
\]
is the cumulative logit for $Y$. While \eqref{eq:lrw} is a standard logistic model,  \eqref{eq:lry} is a proportional odds model, see \cite{Kateri2014}$^7$, Ch. 8, for more  details. Notice that we allow for the interaction between $X$ and $M$ in the outcome equation.  If the assumption of proportionality is not met, the coefficients in the outcome regression may vary with $j$, as in \cite{VdWVanZhang2016}$^3$. We do not pursue this further here, as it is a trivial extension of our proposed derivations. 
%\st{Notice further that }
%\[{\color{red}
%\logit P(Y>j\mid X=x,M=m,C=c) = -\alpha_j + (\beta^{X}x+\beta^{M}m+\beta^{XM}xm+ \beta^Cc), \,\, j=1,\ldots, J-1}
%\]
%\st{\noindent which shows that, if all coefficients are positive, an increase in the value of $X$ increases the probability that $Y$ exceeds the threshold $j$. }
Let $D_j=I(Y\leq j)$  be an indicator variable that takes value 1 if $Y\leq j$ and 0 otherwise. It then follows (see eAppendix A.2 for details) that the marginal cumulative logit model of $Y$ against $X$ can be so written

\begin{equation}\label{eq:margy}
\logit P(Y\leq j\mid X=x, C=c) = \alpha_j-(\beta^Xx +\beta^Cc)-\mbox{log}\left[\frac{1+\mbox{exp}\, g_0^j(x;c)}{1+\mbox{exp} \,g_1^j(x;c)}\right],\,\,\, j=1,\ldots, J-1
\end{equation}
\noindent where the term in squared brackets
 is the relative risk of $\bar M=1-M$ for varying $D_j$ in the distribution of $X=x$ and $C=c$. 
The parametric expression  of the functions $g_d^j$, $d=0,1$, is in (11)
%\eqref{eq:gyx1} 
of eAppendix A.2.

In what follows all effects should be interpreted as conditional on covariates $C=c$. For each level $j$ of the outcome $Y$, the total causal effect involves a contrast between the marginal model evaluated at two different values of $X$, i.e. $x$ and a baseline value $x^*$. We then have
\[\log TCE^j=\beta^{X}(x-x^*)-\mbox{log}\left[\frac{1+\mbox{exp}\, g_1^j(x;c)}{1+\mbox{exp} \,g_0^j(x;c)}\right]+ \mbox{log}\left[\frac{1+\mbox{exp}\, g_1^j(x^*;c)}{1+\mbox{exp} \,g_0^j(x^*;c)}\right]
\] 
\noindent while
\[
\log CDE^j(m) =  (\beta^X + \beta^{XM}m) (x-x^*).
\]
It follows from the proportional odds assumption that the $CDE$ does not vary with the level $j$. 

%\section{Exact parametric formulation of natural effects}

As shown in \cite{VdWVanZhang2016}$^3$, the identification assumptions imply that 
\begin{eqnarray}\nonumber
\logit P(Y(x, M(x^*)) \leq j \mid C=c)= \hspace{8cm}\\[1ex]
\log \frac{\sum_m P(Y\leq j | X=x,M=m,C=c)P(M=m \mid X=x^*, C=c)}{\sum_m P(Y> j |X=x,M=m,C=c)P(M=m \mid X=x^*, C=c)}. \label{vdv}
\end{eqnarray}

The parametric expression of the natural effects can be derived by plugging into \eqref{vdv} the probabilities as implied from model \eqref{eq:lrw} and \eqref{eq:lry}.  Let
$g_{d}^j(x, x^*;c)$ as in (13) %\eqref{eq:gyx2} 
of eAppendix A.2.  After some derivations, that closely resemble the work in \cite{doretti2018exact}$^4$ for the binary-binary case, it is possible to show that 
\[
\log NDE^j=  \beta^{X}(x-x^*)-  \mbox{log}\left[\frac{1+\mbox{exp}\, g_1^j(x,x^*;c)}{1+\mbox{exp} \,g_0^j(x,x^*;c)}\right] + \mbox{log}\left[\frac{1+\mbox{exp}\, g_1^j(x^*;c)}{1+\mbox{exp} \,g_0^j(x^*;c)}\right] 
\]
while 
\[
\log NIE^j= -\mbox{log}\left[\frac{1+\mbox{exp}\, g_1^j(x;c)}{1+\mbox{exp} \,g_0^j(x;c)}\right]+ \mbox{log}\left[\frac{1+\mbox{exp}\, g_1^j(x,x^*;c)}{1+\mbox{exp} \,g_0^j(x,x^*;c)}\right]. 
\]
The above expressions can be further simplified, after their parametric formulation is made explicit. 

%\section{A coherent model for the counterfactual probability}

Notice that the previous derivations allow formulation of the counterfactual model of $Y(x,M(x^*))$, in a way that is coherent with the parametric expressions of the natural direct and indirect effects above introduced. In fact,  $g_{d}^j(x, x^*;c)$ is obtained after crossing the conditional distribution of $Y$ given $M=m$ and $X=x$ with the conditional distribution of $M$ given $X=x^*$, that is

\[g_{d}^j(x, x^*;c)= \log\frac{P(D_j=d\mid M=1,X=x,C=c)}{P(D_j=d\mid M=0,X=x,C=c)} + \log\frac{P(M=1\mid X=x^*,C=c)}{P(M=0\mid X=x^*,C=c)} \]

\noindent with $g_{d}^j(x, x;c)=g_{d}^j(x;c)$. It then follows that the counterfactual model is
\begin{equation}\label{eq:margy}
\logit P(Y(x, M(x^*)) \leq j \mid C=c) = \alpha_j-(\beta^Xx +\beta^Cc)-  \mbox{log} \left[\frac{1+\mbox{exp}\, g_0^j(x,x^*;c)}{1+\mbox{exp} \,g_1^j(x,x^*;c)}\right] ,
\end{equation}
for $j=1,\ldots, J-1$.
This equation, which can be seen as a natural effect model   as in \cite{Langeetal2021}$^8$, shows that  the counterfactual cumulative logit model implied by the postulated data-generating mechanism is a complex function of $x$ and $x^*$ and of the parameters of \eqref{eq:lrw} and \eqref{eq:lry}. 
%\st{ The equation may constitute the basis for estimation methods based on imputations when multiple mediators are present; see Vansteelandtetal2012. Notice that} By inversion,  the effects measured on difference or ratios of probabilities can be easily derived. 

An insight to the causal effects estimates,  their precision and the effect of sparsity is gained via simulation studies that are presented in eAppendix B.1. %\ref{simulations}. The simulations show a good degree of accuracy even when the response is unbalanced (s. Tables 3 and 4 in eAppendix B.1).
The corresponding  bootstrap standard deviations and percentile bootstrap 95\% confidence intervals (CIs) for the causal effect measures are also provided. All the associated {\ttfamily R}-code is provided in eAppendix B.2. %\ref{R-code}.

\section*{Discussion}

This paper extends previous work on parametric mediation analysis to cover a situation with a binary mediator and an ordinal outcome, by deriving exact formulation of the causal effects on the log odds ratio scale. The formulation of the natural effect model for the postulated data-generating mechanism is also derived, thereby allowing the estimation of effects on a different scale. The proposal makes use of well-known statistical models, such as the logistic and the cumulative ordered model, both of them widely used in epidemiologic studies. The methodology  inherits all advantages and limitations of the context. In particular, we here stress the importance of sensitivity analysis to assess that the identification conditions are met in order to make valid causal statements,   see \cite{VanderWeele2015book}$^2$, Ch. 3.

\section*{Acknowledgments}
Elena Stanghellini gratefully acknowledges D.A.A.D. for a research grant (Funding Programme Num. 57552335) that supported her visit to Maria Kateri, during which this paper was developed.

%\end{document}
\newpage

%------------------ Figure ------------------------------

%------------------ APPENDICES ------------------------------
\newpage
\section*{Exact mediation analysis for ordinal outcome and binary mediator  -- eAppendix}
\section*{eAppendix A.1: Causal effects for an ordinal outcome}

The definitions of the causal effects for an ordinal outcome on a logit scale are based on the work of \cite{VdWVanZhang2016}$^3$. Conditional on covariates $C=c$, for each level $j$ of $Y$,  the total causal effect is defined as
\begin{equation}
TCE^j=\frac{P(Y(x)>j \mid  C=c)}{P(Y(x)\leq j\mid C=c)}\left/ \frac{P(Y(x^*)>j\mid C=c)}{P(Y(x^*)\leq j\mid C=c)}\right.
\end{equation}
and measures the extent to which the change in the exposure level from $x^*$ to $x$ increases (or decreases) the odds that the outcome exceeds the level $j$. 

Likewise, the controlled direct effect is defined as
\begin{equation}\label{eq:cdedef}
CDE^j(m) = \frac{P(Y(x,m)>j\mid C=c)}{P(Y(x,m)\leq j\mid C=c)}\left/\frac{P(Y(x^*,m)>j\mid C= c)}{P(Y(x^*,m)\leq j\mid C=c)}\right.
\end{equation}
and describes the causal effect of the exposure on the outcome not mediated by $M$ which is fixed to level $m\in\{0,1\}$. In the other hand, the natural direct effect is obtained as
\begin{equation}\label{eq:ndedef}
NDE^j = \frac{P(Y(x,M(x^*))>j\mid C=c)}{P(Y(x,M(x^*))\leq j \mid C=c)}\left/\frac{P(Y(x^*,M(x^*))>j\mid C=c)}{P(Y(x^*,M(x^*))\leq j\mid C=c)}\right.
\end{equation}
and quantifies the same effect when keeping the mediator to $M(x^*)$, that is, to the level it would have naturally attained under the exposure level $x^*$. The natural indirect effect is given by
\begin{equation}\label{eq:niedef}
NIE^j = \frac{P(Y(x,M(x))>j\mid C=c)}{P(Y(x,M(x))\leq j\mid C=c)}\left/\frac{P(Y(x,M(x^*))>j \mid C=c)}{P(Y(x,M(x^*)\leq j\mid C=c)}\right.
\end{equation}
and compares the odds that $Y$ exceeds the level $j$ had the exposure been set to $x$ and the mediator been set to the value that it would  have naturally attained if the exposure had been set to $x$, i.e. $M(x)$, against the same odds had the exposure been set to $x$ but the mediator been set to the value that it would  have naturally attained if exposure had been set to $x^{*}$, i.e. $M(x^*)$.  Notice that

$$TCE^j=NDE^j \times NIE^j.$$

A similar decomposition of the total causal effect is obtained after interchanging the role  of $x$ and $x^*$ for the potential outcome of $M$; see \cite{RobinsGreen1992}$^9$. 

As $x^*$ is usually considered the baseline category, the difference concerns the apportioning to the direct or indirect effect of a possible interaction between $X$ and $M$ on the outcome $Y$ in the log odds scale. The choice depends on the data at hands and on subject matter considerations. We here refer to this definition, as the other can be derived in analogous way. 

In order to translate the counterfactual entities into something observable, the consistency and composition assumptions are needed, see \cite{Cole and Frangakis 2009}$^{10}$, \cite{vanderweele2009}$^{11}$ and \cite{pearl2010}$^{12}$. The former postulates that for units with exposure level set to $X=x$, the counterfactual value of $M(x)$ coincides with the observed one and, also, for units with $X=x$ and $M=m$, the counterfactual value of $Y(x,m)$ coincides with the observed one. The latter postulates that for units with exposure level set to $X=x$, the counterfactual value $Y(x)$ equals the counterfactual value letting $M$ free to take its counterfactual value $M(x)$, i.e. $Y(x)=Y(x,M(x))$. 

 Furthermore, different conditions based on the conditional independence notion are necessary to identify the causal effects, as detailed in \cite{VdWVanZhang2016}$^3$. In what follows we shall use the notation $A \ind B \mid C$ to indicate that $A$ is independent of $B$ given $C$.  The total causal effect is identified if $Y(x) \ind X \mid C$, i.e.  there is no unobserved confounder of the exposure-outcome relationship. The controlled direct effect is identified if:
$$ Y(x,m) \ind X \mid C \,\mbox{for all} \,x, m \,\,(A.1)$$
$$Y(x,m) \ind M \mid X, C \,\mbox{for all} \,x, m \,\, (A.2)$$

\noindent i.e. there is no unobserved confounder of both the exposure-outcome $(A.1)$ and mediator-outcome $(A.2)$ relationship. In order to identify the natural causal effects, in addition to assumptions $(A.1)$ and $(A.2)$, it is also necessary that
$$M(x) \ind X \mid C \, \mbox{for all} x \,\,(A.3)$$
$$Y(x,m) \ind M(x^*) \mid C \, \mbox{for all} x, x^*, m \,\, (A.4)$$

\noindent i.e. there is no unobserved confounder of exposure-mediator relationship $(A.3)$ and of the mediator-outcome relationship across the two worlds, one where the outcome is free to vary as if the exposure is set to $x$ and the other where the mediator is free to vary as if the exposure is set to $x^*$ $(A.4)$. This last assumption is also known as cross-world independence. For a discussion on the identification assumptions see \cite{vanderweele2009b}$^{13}$, \cite{ShpitserVanderWeele2011}$^{14}$, \cite{steen2018mediation}$^{15}$ and \cite{Didelez2020}$^{16}$.

\section*{eAppendix A.2: Derivation of the exact parametric formulation of the effects}

The interest is in the marginal model of $Y$ against $X$, as a function of the parameters in \mbox{\eqref{eq:lry}} and \mbox{\eqref{eq:lrw}}. From first principles of probability, it follows that:
\begin{equation}\label{eq:wkfy}
\begin{split}
\log\frac{P( D_j=1\mid X=x, C=c)}{P( D_j=0 \mid X=x,C=c)} = & \\
=-\log\frac{P(M=m\mid  D_j=1,X=x,C=c)}{P(M=m\mid  D_j=0,X=x,C=c)} + \log\frac{P( D_j=1\mid M=m,X=x,C=c)}{P( D_j=0\mid M=m,X=x, C=c)},  
\end{split} 
\end{equation}
for $j=1,\ldots, J-1$. The second term of the right hand side of the above equality is given from model \mbox{\eqref{eq:lry}}, while the parametric expression of the first term is not immediately derived from models \mbox{\eqref{eq:lrw}} and \mbox{\eqref{eq:lry}}. 
However, by repeated use of the previous relationship, we have
\begin{equation}\label{eq:wkfw}
\begin{split} 
g_{d}^j(x;c)=\log\frac{P(M=1\mid D_j=d,X=x,C=c)}{P(M=0\mid D_j=d,X=x,C=c)} = &\\
 \log\frac{P(D_j=d\mid M=1,X=x,C=c)}{P(D_j=d\mid M=0,X=x,C=c)} + \log\frac{P(M=1\mid X=x,C=c)}{P(M=0\mid X=x,C=c)},
\end{split}
\nonumber
\end{equation}
 for $d=0,1$ and $j=1,\ldots, J-1$. Using \mbox{\eqref{eq:lry}} and \mbox{\eqref{eq:lrw}}, after some simplifications, we find:
\begin{equation}\label{eq:gyx1}
g_{d}^j(x;c)= -d(\beta^{M}+\beta^{XM}x) +\log\frac{1+\exp (\alpha_{j}-\beta^{X}x-\beta^C c)}{1+\exp(\alpha_j-\beta^{X}x-\beta^{M}-\beta^{XM}x-\beta^C c)} + \gamma_{0}+\gamma_{x}x+\gamma^Cc,
\end{equation} for $d=0,1$ and $j=1,\ldots, J-1$.
%\noindent 
Notice that $g_{d}^j(x;c)$ depends on $j$ only through $\alpha_j$.  Since $[1+\exp g_{d}^j(x;c)]^{-1}$ corresponds to $P(M=0 \mid D_j=d, X=x, C=c)$, substituting in \mbox{\eqref{eq:wkfy}} for $m=0$, we find:

\begin{equation}\label{eq:wkfymar}
\log\frac{P(Y\leq j\mid X=x, C=c)}{P(Y>j\mid X=x, C=c)} =  \alpha_j -\beta^{X}x - \beta^C c - \log {RR}_{\bar M\mid D_j, X=x,C=c},
\end{equation}
where
\[
RR_{\bar M\mid D_j, X=x,C=c}=\frac{1+\exp g_{0}^j(x;c) }{1+\exp g_{1}^j(x;c) }
\]
is the relative risk of $\bar M=1-M$ for varying $D_j$ in the distribution of $X=x$ and $C=c$,  i.e.
\[
RR_{\bar M\mid D_j, X=x, C=c}=  \frac{P(M=0\mid D_j=1, X=x,C=c)}{P(M=0\mid D_j=0, X=x,C=c)}.
\]

\noindent Addition of interaction terms between $C$ and $X$ in the mediator equation and between $C$, $X$ and $M$ in the outcome equation can be done in  straightforward manner; see \cite{stanghellini2019marginal}$^{17}$ for the details with reference to the non counterfactual framework.

The function $g_{d}^j(x;c)$ can be augmented by an argument $x^*$. It therefore becomes:
\begin{equation}\label{eq:gyx2}
g_{d}^j(x,x^*;c)= -d(\beta^{M}+\beta^{XM}x) +\log\frac{1+\exp (\alpha_{j}-\beta^{X}x-\beta^Cc)}{1+\exp(\alpha_j-\beta^{X}x-\beta^{M}-\beta^{XM}x-\beta^Cc)} + \gamma_{0}+\gamma_{x}x^*+\gamma^Cc.
\end{equation}

\section*{eAppendix B.1: Simulation studies and example}\label{simulations}

We conducted two simulation studies, both considering an ordinal response $Y$, a binary mediator $M$ and a continuous 
explanatory variable $X$. In both studies we generated 1000 datasets of size $n=500$. They differentiate in terms of the
number of levels of the response variable. In the first study, the response has $J=3$ levels while in the second $J=5$.
In both cases, variable $M$ was simulated based on model (\ref{eq:lrw}) with parameter values $\gamma_0=-1.0$
and $\gamma^X=0.5$, while $X$ was simulated by a normal distribution with $\mu=3$ and $\sigma=1.5$.
Variable $Y$ was simulated based on (\ref{eq:lry}) with parameters 
$\beta^X=1.1$, $\beta^M=0.7$, $\beta^{XM}=0.5$ and $\alpha=(2.5, 5.5)$, for $J=3$, and 
$\beta^X=0.5$, $\beta^M=1.3$, $\beta^{XM}=0.6$ and $\alpha=(0.5, 2.5, 4.5, 5.5)$, for $J=5$.  All effects here are evaluated at $x^*=2$ and $x=3.5$. This corresponds to an increase of one standard deviation around central values of the treatment. 

The true values of the causal effects, their estimates based on the corresponding mean values of the simulated datasets and the associated asymptotic 95\% confidence intervals (CIs) are provided in Tables \ref{table1} and \ref{table2}, 
for the $J=3$ and $J=5$ cases, respectively. 

\begin{table}[h]
\caption{True values of the causal effects (in the log odds ratio scale) for the first simulation study setup with $J=3$, along with the means and standard deviations of their estimates based on 1000 Monte Carlo (MC) simulated samples of size $n=500$, and the associated 95\% CI for their expected values.}\label{table1}
%
%%% Table for the case J=3:

{\small \begin{center}
\begin{tabular}{llll}
\hline\hline
Effect &  & \multicolumn{2}{c}{Level $j$ of response $Y$}  \\
& & 1 & 2  \\
\hline
log NDE & true value &  1.588 & 1.878  \\
    & MC mean (sd) & 1.612 (0.157) & 1.901 (0.160) \\
		& 95\% CI & (1.313, 1.943) & (1.610, 2.234) \\
\hline
log NIE & true value &  0.378 & 0.381   \\
    & MC mean (sd) &  0.379 (0.060) & 0.384 (0.064) \\
		& 95\% CI & (0.261, 0.505) & (0.260, 0.517) \\
\hline
log TCE & true value &  1.966 & 2.259 \\
    & MC mean (sd) & 1.991 ( 0.159) & 2.285 (0.173) \\
		& 95\% CI & (1.693, 2.313) & (1.976, 2.666)\\
\hline
& & $m=1$ & $m=0$  \\
\hline
log CDE & true value &  2.40 & 1.65   \\
    & MC mean (sd) &  2.435 (0.222) & 1.679 (0.214)  \\
		& 95\% CI & (2.037, 2.903) & (1.281, 2.120) \\
\hline
\hline
\end{tabular}
\end{center}}
\end{table}

%---- Tables for the case J=5:

\begin{table}[h]
\caption{True values of the causal effects (in the log odds ratio scale) for the second simulation study setup with $J=5$, along with the means and standard deviations of their estimates based on 1000 Monte Carlo (MC) simulated samples of size $n=500$, and the associated 95\% CI for their expected values.}\label{table2}
{\small \begin{center}
\begin{tabular}{llllll}
\hline\hline
Effect &  & \multicolumn{4}{c}{Level $j$ of response $Y$}  \\
& & 1 & 2 & 3 & 4 \\
\hline
log NDE & true value &  0.720 & 0.695 & 1.160 & 1.388 \\
    & MC mean (sd) & 0.736 (0.128) & 0.703 (0.090) & 1.172 (0.119)&  1.402 (0.138)  \\
		& 95\% CI & (0.497, 1.005) & (0.540, 0.878) & (0.965, 1.411) & (1.155, 1.681) \\
\hline
log NIE & true value &  0.441 & 0.511 & 0.443 & 0.372  \\
    & MC mean (sd) & 0.441 (0.064) & 0.511 (0.076) & 0.445 (0.068) & 0.373 (0.059) \\
		& 95\% CI & (0.316, 0.578) & (0.365, 0.668) & (0.318, 0.584) & (0.263, 0.492) \\
\hline
log TCE & true value &  1.161 & 1.205 & 1.603 & 1.760 \\
    & MC mean (sd) & 1.177 ( 0.135) & 1.214 (0.103) & 1.617 (0.130) & 1.775 (0.146) \\
		& 95\% CI & (0.942, 1.461) & (1.023, 1.431) & (1.387, 1.889) & (1.511, 2.075) \\
\hline
& & $m=1$ & $m=0$ & & \\
\hline
log CDE & true value &  1.65 & 0.75 &  &  \\
    & MC mean (sd) &  1.665 (0.154) & 0.768 (0.157)  &  & \\
		& 95\% CI & (1.388, 1.978) & (0.461, 1.100) &  &\\
\hline
\hline
\end{tabular}
\end{center}}
\end{table}

\begin{figure}[h]
\centering
 \includegraphics[scale=0.4]{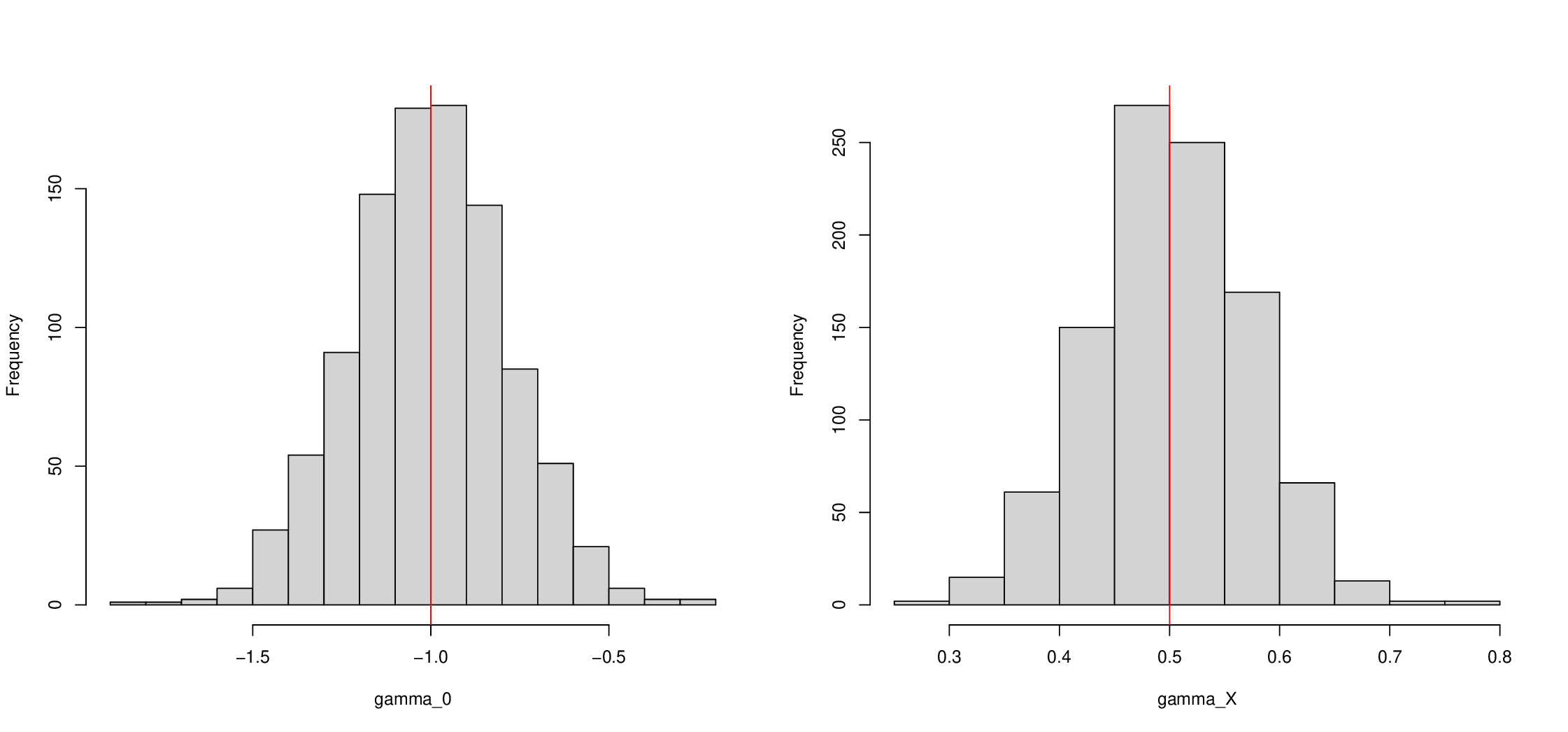} \\
\caption {The histogram of the parameter estimates of the binary logistic mediation model for the 1000 simulated datasets with $J=3$ and $n=500$.}
\label{fig-M}
\end{figure}

\begin{figure}[h]
\centering
 \includegraphics[scale=0.4]{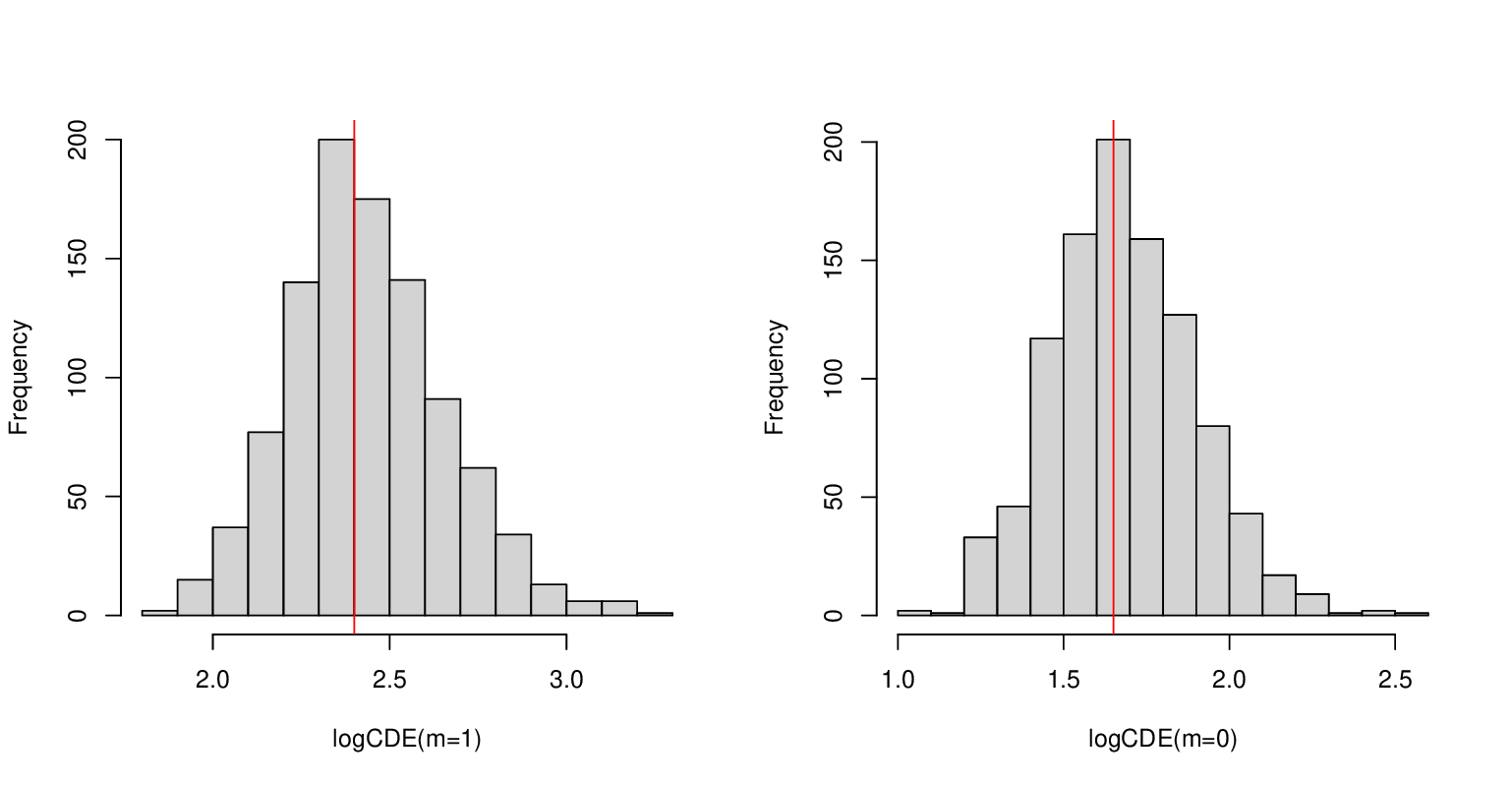} \\
\caption {The histogram of the log(CDE) estimates (for $m=1$ and $m=0$) for the 1000 simulated datasets with $J=3$  and $n=500$.}
\label{fig-CDE}
\end{figure}

For the simulated datasets having a response variable with $J=3$, we do also provide histograms of the estimates of the causal
effects (in log odds ratio scale), the  $\log CDE$ estimates, the parameter estimates for model (\ref{eq:lrw}) and the parameter estimates
for model (\ref{eq:lry}), in Figures \ref{fig-M} up to \ref{fig-CDE}, respectively. All simulations show a good degree of precision of the causal effect estimates obtained. In both scenarios, the $\log{NDE}$ tends to be slightly overestimated, as a possible consequence of the tendency of $\beta_X$ to be overestimated. 
%---- all plots for the case J=3 (n=500) ---------

\begin{figure}[h]
\centering
 \includegraphics[scale=0.35]{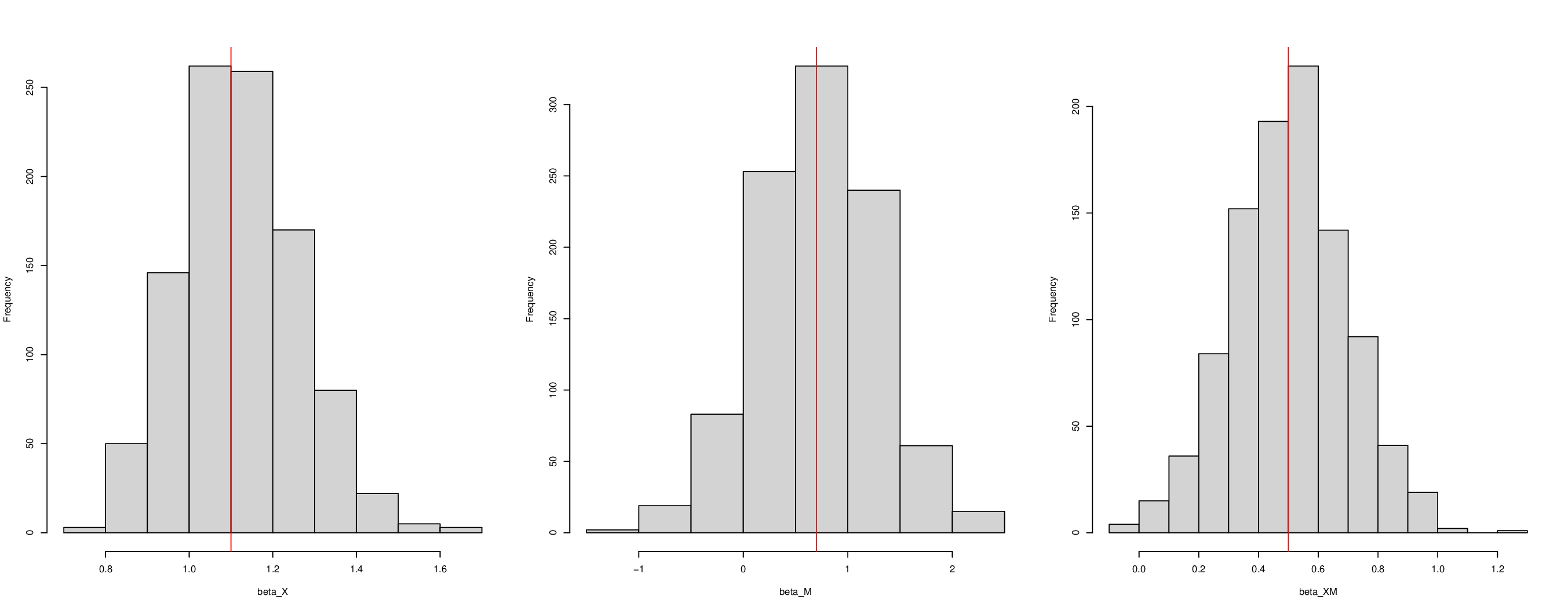} \\
 \includegraphics[scale=0.35]{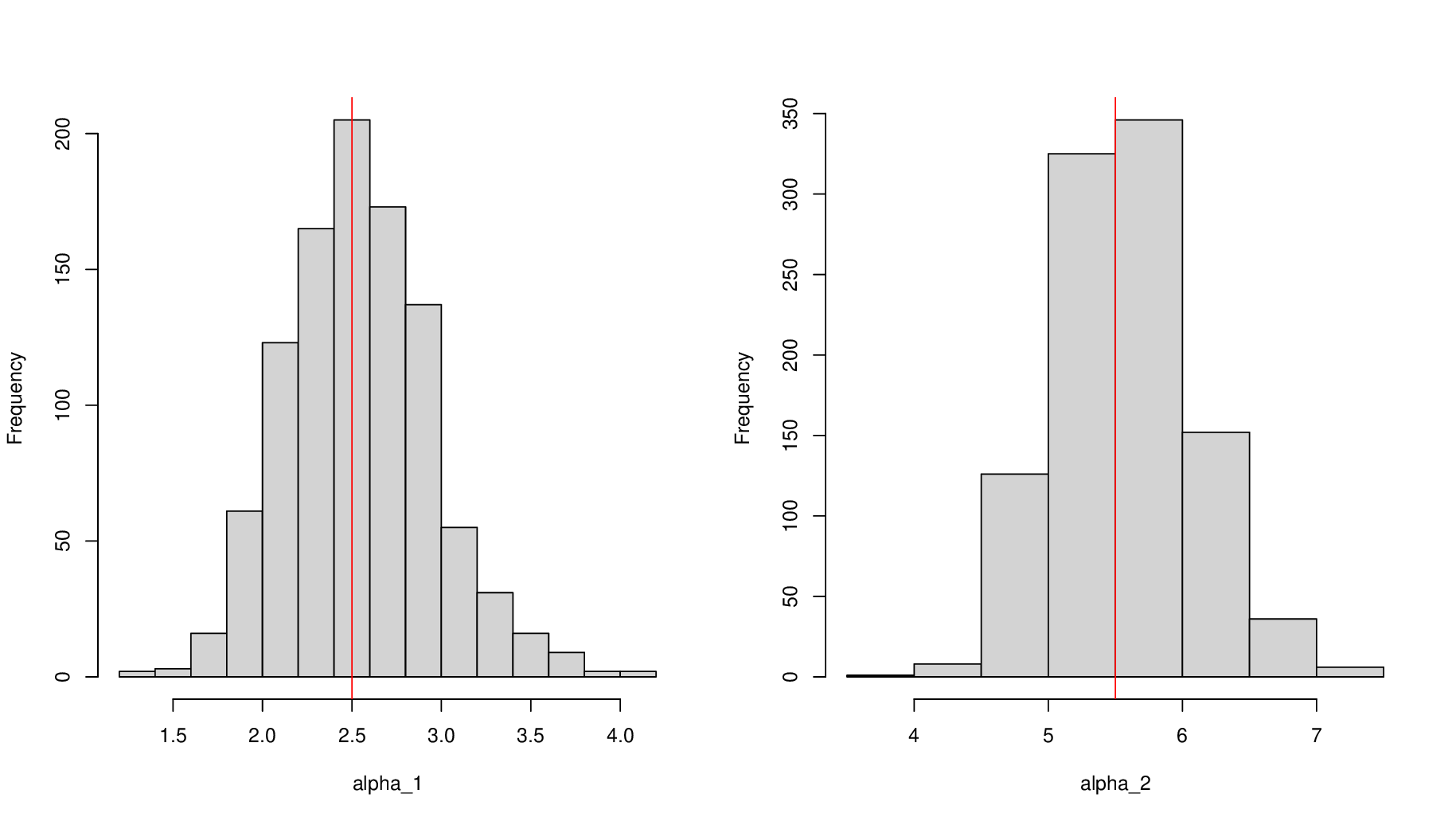} 
\caption {The histogram of the parameter estimates of the ordinal response cumulative logit model for the 1000 simulated datasets with $J=3$ and $n=500$.}
\label{fig-Y}
\end{figure}

\begin{figure}[h]
\centering
 \includegraphics[scale=0.7]{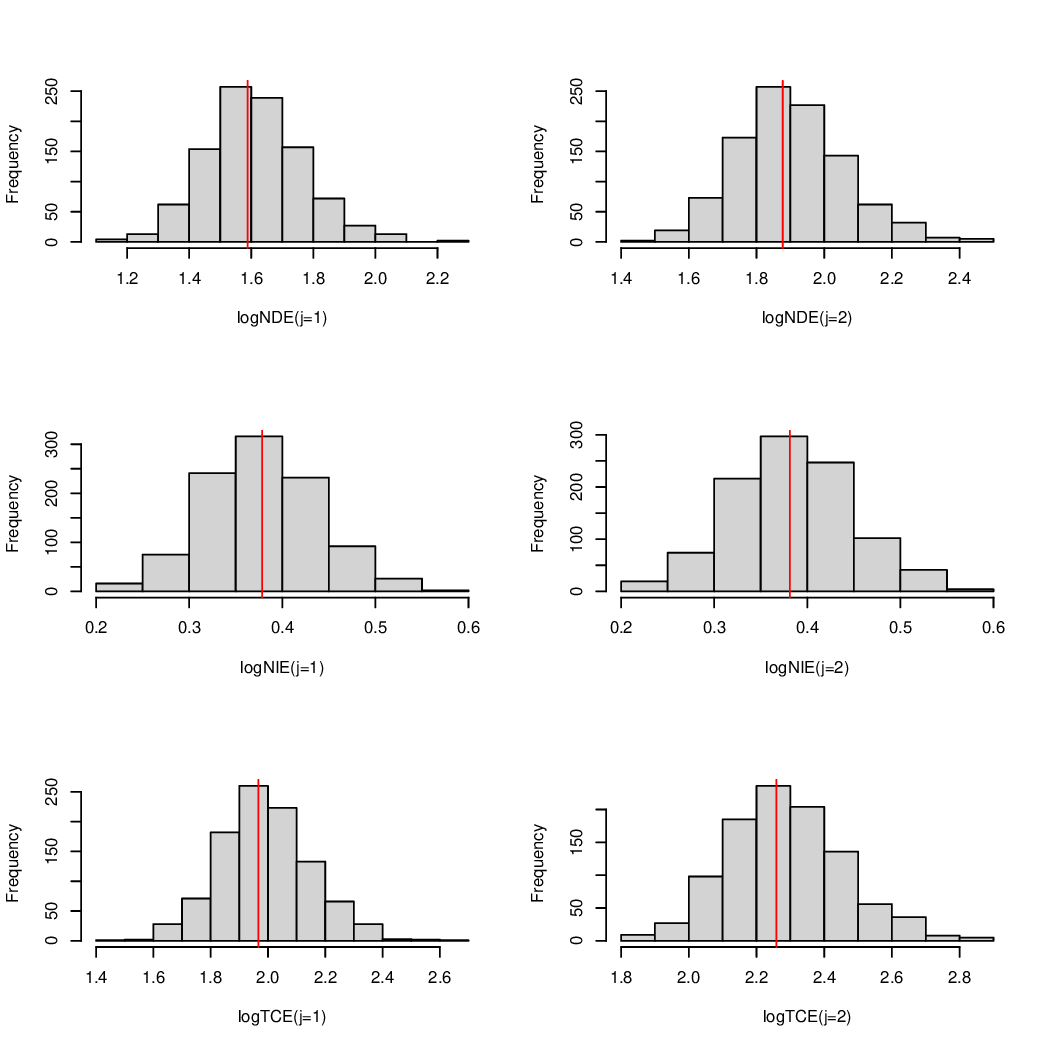} \\
\caption {The histogram of the causal effects estimates (in the log odds ratio scale) for the 1000 simulated datasets with $J=3$  and $n=500$.}
\label{fig-CE}
\end{figure}

In order to gain a view on the effect of sparsity, we simulated a third dataset with an ordinal response $Y$ of $J=5$ levels, a binary mediator $M$ and a continuous explanatory variable $X$. We consider a smaller sample size $n=300$ and set the parameter
values for models (\ref{eq:lrw}) and (\ref{eq:lry}) equal to
$\gamma_0=-1.0$, $\gamma^X=0.9$, $\beta^X=0.5$, $\beta^M=1.3$, $\beta^{XM}=0.6$ and $\alpha=(-0.9, 0.9, 2.2, 3.5)$. $X$
is again simulated by a normal distribution with $\mu=3$ and $\sigma=1.3$.  In Table \ref{tablemarg} we report the marginal table according to $Y$. Notice that the last category contains 
59\% of units, a rather different  scenario from the one considered in \cite{VdWVanZhang2016}$^3$, also in that here $M$ is binary. 

\begin{table}[ht]
\centering
\caption{Observed frequencies according to $Y$ in the sparse dataset}\label{tablemarg}
\begin{tabular}{rrrrr} \hline\hline
\multicolumn{5}{c}{Level $j$ of response $Y$} \\
1 &  2 &  3 &  4 & 5 \\ \hline
8 &  27 &  39 & 49 & 177 \\ 
  \hline\hline
\end{tabular}
\end{table}

We fitted models 
(\ref{eq:lrw}) and (\ref{eq:lry}) and obtained the following parameter estimates:
$\hat\gamma_0=-1.0005$, $\hat\gamma^X=0.8228$, $\hat\beta^X=0.5344$, $\hat\beta^M=0.8088$, $\hat\beta^{XM}=0.6939$
and $(\hat\alpha_1,\hat\alpha_2,\hat\alpha_3,\hat\alpha_4)=(-1.2887, 0.7009, 2.1666, 3.5760)$.
For these parameter values, we simulated in the sequel 100 bootstrap datasets and computed the causal effect values for $x^*=2$ and $x=3.5$. 
The true value and the corresponding 95\% percentile bootstrap CIs are in Table \ref{table3}. Despite sparsity, results show a rather good precision of the estimates, with a small tendency of $\log NDE$ to be overestimated and of $\log NIE$ to be underestimated.
\begin{table}[h]
\caption{Estimated causal effects values (in the log odds ratio scale) for the sparse data example, corresponding bootstrap standard deviations and percentile bootstrap 95\% CIs, 
based on a 1000 bootstrapped samples. The example is of sample size $n=300$. The true causal effects values are also provided. }\label{table3}
{\small \begin{center}
\begin{tabular}{llllll}
\hline\hline
Effect &  & \multicolumn{4}{c}{Level $j$ of response $Y$}  \\
& & 1 & 2 & 3 & 4 \\
\hline
log NDE & true value & 0.819 & 0.751 & 0.781 & 1.048 \\
 & estimate & 0.906 & 0.855 &  0.892 & 1.205 \\
& boot.sd &  0.257 & 0.226 & 0.176 & 0.171 \\
		& 95\% boot.CI & (0.411, 1.410) & (0.455, 1.342) & (0.600, 1.267) & (0.923, 1.585)\\
\hline
log NIE & true value &  0.912 & 0.913 & 0.864 & 0.681  \\
 & estimate &  0.801 & 0.811 & 0.790 & 0.640  \\
& boot.sd &  0.151 & 0.154 & 0.152 & 0.120 \\
		& 95\% boot.CI & (0.530, 1.132) & (0.538, 1.153) & (0.520, 1.134) & (0.422, 0.902)\\
\hline
log TCE & true value &  1.730 & 1.664 & 1.645 & 1.729 \\
 & estimate &   1.707 & 1.666 & 1.683 & 1.845 \\
& boot.sd &  0.261 & 0.230 & 0.193 & 0.210 \\
		& 95\% boot.CI & (1.254, 2.273) & (1.283, 2.143) & (1.363, 2.108) & (1.506, 2.316)\\
\hline
& & $m=1$ & $m=0$ & & \\
\hline
log CDE & true value &  1.650 & 0.750 &  &  \\
 & estimate &  1.842 & 0.802 &  &  \\
& boot.sd &  0.249 & 0.280 &  & \\
		& 95\% boot.CI & (1.433, 2.416) & (0.261, 1.371) &  &\\
\hline
\hline
\end{tabular}
\end{center}}
\end{table}

%--------------------------------------------------------

\section*{eAppendix B.2: {Implementation in \ttfamily R}}\label{R-code}

In this appendix we provide the {\ttfamily R} code used for our analysis. The true values of the causal effects for the setup of the second simulation study (with $J=5$) are computed as follows. The {\ttfamily cause} function called is provided in next section.\\

\noindent
{\ttfamily
J <- 5   \hspace{4.64cm}      \# number of levels of Y \\
gam0 <- -1.0;  gamx <- 0.5  \hspace{0.45cm}  \# gamma\_0 and gamma\_x for model (5)  \\
alpha <- c(0.5,2.5,4.5,5.5) \  \# parameters for model (6) for Y  \\
beta\_x= 0.5;  beta\_m= 1.3; beta\_xm= 0.6  \\
meanX=3; sdX=1.5           \ \ \  \ \ \ \ \# X simulated from N(meanX,sdX) 

\noindent \# True values for the causal effects (in log-scale), for $x$ and $x^*$:  \\
x <- 3.5; xstar <- 2         \  \ \  \# values of fixed $x$ and $x^*$  \\
cef <- cause(J,alpha,beta\_x,beta\_m,beta\_xm,gam0,gamx,x,xstar)  \\
NDEtrue <- cef\$NDE; NDEtrue  \\
NIEtrue <- cef\$NIE; NIEtrue  \\
TCEtrue <- cef\$TCE; TCEtrue  \\
CDEtrue <- cef\$CDE; CDEtrue   \\ }

The basic function {\ttfamily simul} used in the simulation studies requires functions of the {\ttfamily MASS} library and is provided in  Section B.3 that follows. %\ref{sec-R-functions}.

The example presented was simulated by the code provided below.\\

\noindent
{\ttfamily
\# Simulate the data set (of size 300): \\
n <- 300 \\
data <- simul(n,meanX,sdX,gam0,gamx,alpha,beta\_x,beta\_m,beta\_xm)\\
table(data\$M)  \# for seeing the table of frequencies for the generated M \\
table(data\$Y)  \# and Y data }\\

The results on our simulated example, given in Table \ref{table3}, are derived in {\ttfamily R} as follows. \\

\noindent
{\ttfamily
\# Estimates of causal effects values (for the x and xstar as above): \\
X <- data\$X; M <- data\$M; Y <- factor(data\$Y)\\
model.M <- glm(M $\sim$ X, family = binomial(link = "logit")); summary(model.M) \\
model.Y<-polr(formula = Y $\sim$ X + M + X:M); summary(model.Y) \\
alpha <- model.Y\$zeta; beta\_x <- model.Y\$coefficients[1] \\
beta\_m <- model.Y\$coefficients[2]; beta\_xm <- model.Y\$coefficients[3]  \\
gam0 <- model.M\$coefficients[1]; gamx <- model.M\$coefficients[2] \\
cef.est <- cause(J,alpha,beta\_x,beta\_m,beta\_xm,gam0,gamx,x,xstar)\\
NDEest <- cef.est\$NDE; \ NDEest \\
NIEest <- cef.est\$NIE; \ NIEest \\
TCEest <- cef.est\$TCE; \ TCEest \\
CDEest <- cef.est\$CDE; \ CDEest

\noindent\# 95\% bootstrap percentile CIs for the mean TCE \\
\# (based on 1000 bootstrap samples): \\
nboot<- 1000 \\
PTCE <- matrix(,nboot,4) \# matrix of nboot rows and 4 columns \\
PNDE <- matrix(,nboot,4); \ \ PNIE <- matrix(,nboot,4) \\
PCDE <- matrix(,nboot,2) \# matrix of nboot rows and 2 columns\\
Psample.ex <- matrix(,n,3)\\
\noindent for(i in 1:nboot)\{ \\
Psample.ex <- example[sample(nrow(example), n, replace=TRUE), ] \\
X <- Psample.ex\$X; M <- Psample.ex\$M; Y <- factor(Psample.ex\$Y) \\
model.M <- glm(M $\sim$ X, family = binomial(link = "logit")) \\
model.Y <- polr(formula = Y $\sim$ X + M + X:M) \\
alpha0 <- model.Y\$zeta; betaX <- model.Y\$coefficients[1] \\
betaM <- model.Y\$coefficients[2]; betaXM <- model.Y\$coefficients[3] \\
gamma0 <- model.M\$coefficients[1]; gammaX <- model.M\$coefficients[2] \\
cef.est <- cause(J,alpha0,betaX,betaM,betaXM,gamma0,gammaX,x,xstar) \\
PTCE[i,] <- cef.est\$TCE; PNDE[i,] <- cef.est\$NDE \\
PNIE[i,] <- cef.est\$NIE; PCDE[i,] <- cef.est\$CDE \ \ \} \\
quantile(PTCE[,1],c(0.025,0.975)); quantile(PTCE[,2],c(0.025,0.975)) \\
quantile(PTCE[,3],c(0.025,0.975)); quantile(PTCE[,4],c(0.025,0.975)) \\
}

The bootstrap CIs for the remaining causal effects are derived analogously using 
the simulated sample values and saved in {\ttfamily PNDE}, {\ttfamily PNIE} and 
{\ttfamily PCDE}, respectively. Alternatively, one could use the {\ttfamily boot} package
with more options for alternative types of bootstrap CIs, like for example the bootstrap BCa CIs.

\section*{eAppendix B.3: {\ttfamily R}-Functions}\label{sec-R-functions}

The {\ttfamily cause} function computes $\log NDE^j $, $\log NIE^j $, $\log TCE^j $, $j=1, \ldots, J-1$, as well as 
 $\log CDE(m)$, $m=0,1$,
for given $x$ and $x^*$ and given parameter values or their estimates for the models (\ref{eq:lrw}) and (\ref{eq:lry}).\\

\noindent
{\small\ttfamily
cause <- function(J,a,bX,bM,bXM,g0,gX,x,xstar)\{ \\
\phantom{.}\hspace{0.2cm}    A <- vector(mode="numeric", length=J-1); Astar <- A; B <- A \\
\phantom{.}\hspace{0.2cm}    fg0<- function(j,x)\{\\
\phantom{.}\hspace{0.5cm}             A <- log((1+exp(a[j]-bX*x))/(1+exp(a[j]-bX*x-(bM+bXM*x))))\\
\phantom{.}\hspace{0.5cm}            return(A+g0+gX*x)\}\\
\phantom{.}\hspace{0.2cm}    fg1<- function(j,x)\{fg0(j,x)-(bM+bXM*x)\}\\
\phantom{.}\hspace{0.2cm}   g <- function(d,j,x,xstar)\{ \# d=1: Prob(Y$\leq$ j) <-> d=0 Prob(Y$>$j) \\
 \phantom{.}\hspace{0.5cm}      fg0(j,x)-(g0+gX*x)+(g0+gX*xstar)+(d==1)*(-(bM+bXM*x))\}\\
\phantom{.}\hspace{0.2cm}     for (j in 1:J-1) \{\\
\phantom{.}\hspace{0.5cm}       A[j] <- (1+exp(fg1(j,x)))/(1+exp(fg0(j,x)))\\
\phantom{.}\hspace{0.5cm}       Astar[j] <- (1+exp(fg1(j,xstar)))/(1+exp(fg0(j,xstar)))\\
\phantom{.}\hspace{0.5cm}       B[j] <- (1+exp(g(1,j,x,xstar)))/(1+exp(g(0,j,x,xstar)))\}\\
\phantom{.}\hspace{0.2cm}    NDE <- exp(bX*(x-xstar))*Astar/B\\
\phantom{.}\hspace{0.2cm}    NIE <- B/A\\
\phantom{.}\hspace{0.2cm}    TCE <- NDE*NIE\\         
\phantom{.}\hspace{0.2cm}    CDE <- c(exp((x-xstar)*(bX+bXM)),exp((x-xstar)*bX))\\
\phantom{.}\hspace{0.2cm}    return(list(TCE=log(TCE),NDE=log(NDE),NIE=log(NIE),CDE=log(CDE)))\}\\
}

The {\ttfamily simul} function next simulates a sample of size $n$
that consists of $X$ values generated by a normal distribution with mean and standard deviation
controlled by {\ttfamily meanX} and {\ttfamily sdX}, and $M$ and $Y$ values generated by models
(\ref{eq:lrw}) and (\ref{eq:lry}), respectively, with corresponding parameters 
{\ttfamily gam0,gamx,alpha,beta\_x,beta\_m,beta\_xm}, where all are scalar except {\ttfamily alpha}
that is a vector of dimension $J-1$.\\

\noindent
{\ttfamily 
simul <- function(n,meanX,sdX,gam0,gamx,alpha,beta\_x,beta\_m,beta\_xm)\{ \\
\phantom{.}\hspace{0.2cm}           J <- length(alpha)+1 \\
\phantom{.}\hspace{0.2cm}           X <- rnorm(n, mean=meanX, sd=sdX) \\
\phantom{.}\hspace{0.2cm}           probM <- exp(gam0+gamx*X)/(1+exp(gam0+gamx*X))  \# Prob(M=1) by model (1)  \\
\phantom{.}\hspace{0.2cm}           M <- rbinom(n,1,probM)  \\
\phantom{.}\hspace{0.2cm}           a0 <- matrix(rep(alpha,n),nrow=J-1)      \\
\phantom{.}\hspace{0.2cm}           bx <- matrix(rep(beta\_x, J-1))\%*\%t(matrix(X))        \\
\phantom{.}\hspace{0.2cm}           bm <- matrix(rep(beta\_m, J-1))\%*\%t(matrix(M))        \\                      
\phantom{.}\hspace{0.2cm}           bxm <- matrix(rep(beta\_xm, J-1))\%*\%t(matrix(X*M))    \\ 
\phantom{.}\hspace{0.2cm}           lcumpY= a0-bx-bm-bxm \hspace{1cm} \# model (6)  \\
\phantom{.}\hspace{0.2cm}           cumpY <- exp(lcumpY)/(1+exp(lcumpY))         \\
\phantom{.}\hspace{0.2cm}           cumpY <- rbind(cumpY,1-cumpY[J-1,])  \\
\phantom{.}\hspace{0.2cm}           zer <- rep(0,n); L <- J-2   \\
\phantom{.}\hspace{0.2cm}           cump0Y <- rbind(zer,cumpY[1:L,],zer)   \\
\phantom{.}\hspace{0.2cm}           py <- cumpY-cump0Y  \hspace{1.2cm}  \# Prob(Y)   \\
\phantom{.}\hspace{0.2cm}           Y <- c();  for (i in 1:n) \{Y[i] <- which(rmultinom(1,1,py[,i])==1)\}  \\
\phantom{.}\hspace{0.2cm}           return(list(X=X,M=M,Y=Y))\}  \\
}

The {\ttfamily R} functions constructed are adjusted to the needs of our simulation study and do not consider
covariates in model (\ref{eq:lrw}) and (\ref{eq:lry}). Their extension to cases with covariates is straightforward.

\end{document}